\documentclass[doublecol,figures]{epl2} 
\usepackage[english]{babel}
\usepackage{amsmath}
\usepackage{float}
\usepackage{bm}
\usepackage{multirow}
\usepackage{dcolumn}
\usepackage{xfrac}

\newcommand{\icm}{\ensuremath{~\textrm{cm}^{-1}}}
\newcommand{\BKFA}{\chem{Ba_{1-x}K_xFe_2As_2}}
\newcommand{\BFCA}{\chem{BaFe_{2-x}Co_xAs_2}}
\newcommand{\BKFAn}{\chem{Ba_{0.6}K_{0.4}Fe_2As_2}}
\newcommand{\BFCAn}{\chem{BaFe_{1.84}Co_{0.08}As_2}}

\title{Optical conductivity of \BKFAn: the effect of in-plane and out-of-plane doping in the the superconducting gap}
\shorttitle{Optical conductivity of \BKFAn: in- and out-of-plane doping} 

\author{Y. M. Dai\inst{1,2} \and B. Xu\inst{2} \and B. Shen\inst{2} \and H. H. Wen\inst{2,3} \and X. G. Qiu\inst{2} \and R. P. S. M. Lobo\inst{1}}
\shortauthor{Y. M. Dai \etal}

\institute{                    
  \inst{1} LPEM, ESPCI-ParisTech, CNRS, UPMC, 10 rue Vauquelin, F-75231 Paris Cedex 5, France\\
  \inst{2} Beijing National Laboratory for Condensed Matter Physics, National Laboratory for Superconductivity, Institute of Physics, Chinese Academy of Sciences, P.O. Box 603, Beijing 100190, China \\
  \inst{3} National Laboratory of Solid State Microstructures and Department of Physics, Nanjing University, Nanjing 210093, China
}

\date{\today}
\abstract{
We measured the \emph{in-plane} optical conductivity of a nearly optimally doped ($T_{c}$ = 39.1 K) single crystal of \BKFAn. Upon entering the superconducting state the optical conductivity vanishes below $\sim$ 20 meV, indicating a fully gapped system.  A quantitative modeling of the optical response of this material requires two different isotropic gaps, the larger of which dominates the London penetration depth. The temperature dependence of these gaps indicate a strong interband interaction, but no impurity scattering induced pair breaking is present. This contrasts to the large residual conductivity observed in optimally doped \BFCAn\ and strongly supports an $s_\pm$ gap symmetry for both compounds.
}
\begin{document}

\maketitle

\bibliographystyle{eplbib}

%
%
\section{Introduction}

A crucial issue in the superconducting mechanism of iron-arsenide compounds is to  understand the properties of their superconducting gaps. Electronic structure calculations predict that multiple bands at the Fermi level participate in the formation of the condensate \cite{PhysRevLett.100.226402, PhysRevLett.100.237003}, leading to multigap superconductivity. As the electron-phonon coupling in these materials is too small to account for $T_c$ \cite{Boeri2008}, an all electronic mechanism such as antiferromagnetic spin fluctuations was suggested for pairing. This leads to a possible sign reversal between the order parameters in different Fermi surface sheets, the so-called $s_{\pm}$ symmetry \cite{PhysRevLett.101.057003, Cvetkovic2009}. In an $s_{\pm}$ superconductor, interband scattering by non-magnetic  impurities is pair-breaking as it mixes hole and electron states having superconducting order parameters of opposite phases \cite{Vorontsov2009,PhysRevB.80.140515,Noat2010}.

In this scenario, BaFe$_2$As$_2$ based materials are of the utmost importance as superconductivity appears by doping either in or out of FeAs planes, which contribute the most to the bands at the Fermi level. Angle resolved photoemission spectroscopy (ARPES) finds multiple nodeless gaps in optimally doped \BFCA\ \cite{Terashima2009} and \BKFA\ \cite{EPL.83.47001,EPL.85.67002} in accordance with theoretical calculations \cite{Maiti2011}. Yet, their spectroscopic responses are strikingly different.

In \BFCA, Co atoms go into the FeAs planes. This material has a V-shaped density of states \cite{PhysRevB.79.054529} that leads to a strong sub-gap absorption \cite{PhysRevLett.106.087004,Gordon2009,PhysRevB.82.100506}. Optical conductivity  \cite{Lucarelli2010,PhysRevB.81.100512,EPL.90.37005,PhysRevB.81.214508,PhysRevB.82.184527,PhysRevB.82.100506,PhysRevB.82.174509,Fischer2010,Aguilar2010} shows multiple superconducting gaps and, most importantly, sub-gap absorption in the superconducting state which cannot be accounted for by thermally broken pairs.
This sub-gap absorption agrees with the absence of a sharp quasiparticle peak in ARPES.

\BKFA\ represents a different situation as doping K atoms sit out of the FeAs planes. Local magnetization \cite{PhysRevLett.101.257006}; point-contact Andreev reflection \cite{PhysRevB.79.012503}; specific heat \cite{PhysRevB.79.174501,PhysRevLett.105.027003}; scanning tunneling microscopy (STM) \cite{PhysRevB.83.060510}; and thermal transport \cite{PhysRevB.80.140503} point towards multiple fully open gaps, in agreement with ARPES. Two superconducting gaps dominate the physics of optimally doped \BKFAn: $\Delta_1 \sim 4$ meV and $\Delta_2 \sim 10$ meV. 

Should we expect any fundamental difference between \BFCA\ and \BKFA\ families? Out-of-plane impurities should not have a major influence on interband scattering, whereas the converse holds for in-plane dopants. 
Indeed, the larger impurity scattering in optimally Co doped BaFe$_2$As$_2$ can be inferred from its residual resistance ratio (RRR) of about 3, against 14 for optimally K doped samples \cite{Rullier2009,Shen2011}.
Can the different location of doping atoms shed any light on the claimed $s_{\pm}$ symmetry? 

Optical conductivity is a probe sensitive to low energy states and provides direct information on unpaired carrier electrodynamics. Opposite to \BFCA, little optical data exists on \BKFA. Yang \etal \cite{PhysRevLett.102.187003} normal state measurements indicate that carriers are coupled to a broad bosonic spectrum extending beyond 100 meV with a large coupling constant at low temperature. Li \etal \cite{PhysRevLett.101.107004} found a fully open gap but could not quantify the presence of multiple superconducting gaps. The Eliashberg calculations by Charnukha \etal \cite{Charnuka2011} describes ellipsometry data with two gaps, having values compatible with other techniques. Dai \etal \cite{Dai2013} showed that the presence of mulitple bands hides a linear temperature dependence of the transport properties.

In this paper, we tackle the problem of the gap symmetry in BaFe$_2$As$_2$ based superconductors. We carry out a detailed infrared study on a nearly optimally doped \BKFAn\ single crystal, where doping K atoms go out of the FeAs plane. In the superconducting phase, we observe a vanishing optical conductivity below 20 meV. However, a second absorption edge at 33 meV is required to quantitatively describe the optical conductivity. This fully open gap indicates the absence of impurity induced pair breaking. This contrasts to optimally doped \BFCAn, where in-plane impurity scattering leads to pair breaking and sub-gap absorption. Our results provide a strong support for an $s_{\pm}$ gap symmetry.

%
%
\section{Experimental Methods}

\begin{figure}[htb]
  \onefigure[width=0.8\columnwidth]{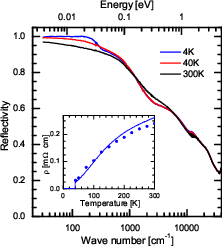}
  \caption{\BKFAn\ infrared-visible-UV reflectivity measured at 4, 40, and 300 K. The inset shows the dc resistivity (solid line) compared to the values from the zero frequency extrapolation of the optical conductivity (solid circles).
}
  \label{Fig1}
\end{figure}

High quality single crystals of \BKFAn\ were grown by an FeAs flux method \cite{Supercond.Sci.Technol.21.125014}. Their resistivity (inset of Fig.~\ref{Fig1}) shows a sharp superconducting transition with an onset at $T_{c}$ = 39.1~K and a width $\Delta T_{c} \sim 0.5$~K. Near normal incidence reflectivity from 20 to $12\,000\icm$ was measured at ESPCI on Bruker IFS113 and IFS66v spectrometers at 19 different temperatures from 4 to 300~K in an ARS Helitran cryostat. The absolute reflectivity ($R$) of the sample was obtained with an \emph{in situ} gold overfilling technique. $R$ has an absolute accuracy better than 0.5\% and a relative accuracy better than 0.1\%. The data was extended to the visible and UV ($10\,000$ to $40\,000\icm$) at room temperature with an AvaSpec-$2048 \times 14$ spectrometer. The sample was cleaved prior to each temperature run. Infrared  measurements were repeated, for a more limited set of temperatures, on the same crystal at IOP with a Bomem DA8 spectrometer. Measurements at ESPCI and IOP agree within 0.1\%. Figure \ref{Fig1} displays the reflectivity of \BKFAn\ on the \emph{ab}-plane above and below $T_c$.

In order to have a fine temperature dependence of the gaps and penetration depth we complemented the reflectivity data with a relative thermal reflectivity (RTR) measurement. In this technique, we determine the reflectivity of the material over a restricted temperature range without any physical motion of the sample. The data at each temperature is divided by the measurement at one arbitrary temperature $T_0$. Next, each temperature is multiplied by the absolute reflectivity measured at $T_0$ using the \textit{in situ} gold overfilling technique. RTR is very useful in determining small changes of the spectrum through a phase transition. Here, we measured the RTR from 5 to 45 K, every Kelvin, in the 50--700 \icm\ spectral range. All spectra were normalized by the measurement at $T_0 = 40$~K. 

%
%
\section{Far-infared reflectivity}
\label{SecIII}

\begin{figure}[htb]
  \onefigure[width=0.8\columnwidth]{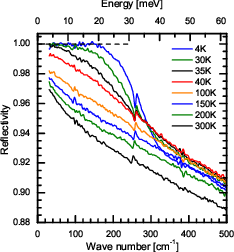}
  \caption{Far-infrared data showing the formation of a sharp reflectivity edge below $T_c$ with a flat reflectivity at low frequencies at 4 K.
}
  \label{Fig2}
\end{figure}

Figure~\ref{Fig2} shows the far-infrared reflectivity of \BKFAn. Above $T_c$ the reflectivity smoothly decreases with frequency, except for the phonon at 260 \icm\ (32 meV) \cite{Dai2012b}. Below $T_c$ a sharp edge develops and the reflectivity rises to a flat 100\% value below 160 \icm\ (20 meV) at the lowest measured temperature, indicating a fully open superconducting gap. 

The accuracy of the 100\% value is an important issue and deserves some remarks. Once we get closer to 100 \%, small errors in the reflectivity are amplified when calculating the real part of the optical conductivity ($\sigma_1$). As we show below, in \BKFAn, a 100 \% reflectivity leads to a fully gapped $\sigma_1$. Considering our claimed absolute accuracy of 0.5\% , a reflectivity of 99.5 \% leads to a low energy residual $\sigma_1$ in the $1000~\Omega^{-1} \textrm{cm}^{-1}$ range. 

A reflectivity of 100 \% is not the only indication of a full gap. The energy dispersion of the data also leads to this conclusion. Our \BKFAn\ sample, at 4~K, has a constant reflectivity below $\sim 160 \icm$. The statistic average value of $R$ at 4 K, between 20 and 160\icm, is $0.9996 \pm 0.00098$.  Was the system not fully gapped, one would have to consider the presence of unpaired quasiparticles that leads to a non vanishing curvature for the reflectivity. This curvature is positive for normal Drude-like carriers and negative for a gap with residual thermally broken quasiparticles. At 4 K, our sample has a flat, constant reflectivity below 160\icm. For comparison, in the 30 and 35 K data, although the reflectivity reaches for 100 \% at low frequencies, a clear frequency dispersion is present at all energies. Of course, at these two temperatures thermally broken pairs lead to a sub-gap absorption. The flat, dispersionless, reflectivity at 4 K is not compatible with unpaired quasiparticles and the only physical interpretation for the data is a fully gapped system.

%
%
\section{Superconducting gaps}
\label{SecIV}

\begin{figure}[htb]
  \onefigure[width=7cm]{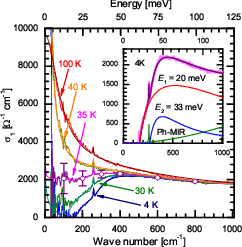}
  \caption{\BKFAn\ optical conductivity above and below $T_c$. The smooth lines through the data are fits assuming that mobile charges are described by either Drude (normal) or Mattis-Bardeen (superconducting) carriers. Inset: decomposition of the optical conductivity fit at 4 K into individual contributions. Note the change of slope in $\sigma_1(\omega)$ at the larger absorption edge, $E_2 = 33$ meV. }
  \label{fig3}
\end{figure}

We derived $\sigma_1(\omega)$ from the reflectivity through Kramers-Kronig analysis. At low frequencies we utilized either a Hagen-Rubens ($1 - A \sqrt{\omega}$) or a superconducting ($1 - A \omega^4$) extrapolation. At high frequencies we applied a constant reflectivity to 40 eV terminated by a $\omega^{-4}$ free electron term. Figure \ref{fig3}  shows $\sigma_1(\omega)$ at various temperatures. Error bars at selected frequencies are shown at 35 K, temperature where they are the largest. The error bars were estimated assuming different high and low frequency extrapolations and an error of 0.5\% in the absolute reflectivity.

Above $T_{c}$, $\sigma_{1}(\omega)$ shows a metallic response modeled by a Drude-Lorentz optical conductivity:
\begin{equation}
\sigma_{1}(\omega)=\frac{2\pi}{Z_0}\sum_k\frac{\Omega^{2}_{p,k}}{\omega^{2} \tau_k +\frac{1}{\tau_k}}+\sum_k\frac{\gamma_k\omega^{2}S_k^{2}}{(\Omega_k^{2}-\omega^{2})^{2}+\gamma_k^{2}\omega^{2}} ,
\label{eq.1}
\end{equation}
where $Z_0$ is the vacuum impedance. The first term corresponds to a sum of free-carrier Drude responses, each having a plasma frequency $\Omega_{p,k}$; and a scattering rate $\tau_k^{-1}$. The second term is a sum of Lorentz oscillators with resonance frequency $\Omega_k$; line width $\gamma_k$; and plasma frequency $S_k$. Band structure theory predicts up to 5 bands at the Fermi level. Nevertheless the optical conductivity is a reciprocal space averaged quantity and, hence, all bands with similar carrier lifetimes will contribute to the same Drude term in $\sigma_1$. In this scenario, our data can be fitted by two Drude terms in agreement with lifetime distributions for pnictides \cite{Analytis2009,Richard2009,Shishido2010,Kemper2011}. Indeed, Wu \etal \cite{PhysRevB.81.100512} showed that two Drude terms are enough to describe the optical conductivity of most iron-arsenide superconductors. We also kept one Lorentz oscillator to represent the phonon at 32 meV and another to describe mid-infrared interband transitions. These four contributions account for the optical conductivity up to 1.25 eV at all measured temperatures in the normal state. Fits obtained for the normal state with Eq.~\ref{eq.1} are shown as thin solid smooth lines in Fig.~\ref{fig3}. We compare the inverse of the zero frequency value of our $\sigma_1(\omega)$ fits to the measured dc resistivity in the inset of Fig.~\ref{Fig1}. The good agreement indicates that our modeling of the data is a reliable parametrization of the system.

Below $T_{c}$, a dramatic suppression of $\sigma_{1}(\omega)$ sets in. At 4~K $\sigma_{1}(\omega)$ vanishes below 20 meV indicating a fully open gap. In order to quantitatively describe the superconducting state optical response, we replaced the two normal state Drude terms in Eq.~\ref{eq.1} by two corresponding Mattis-Bardeen conductivities, modified to take into account arbitrary scattering \cite{PhysicaC.183.99}. The fits thus obtained are also shown as solid smooth lines in Fig.~\ref{fig3}. The inset details the four contributions (two superconducting bands, one phonon and the mid-infrared interband transition) obtained for $\sigma_1(\omega)$ at 4 K. The two Mattis-Bardeen terms have absorption edges at $E^{(0)}_1 = 20$~meV and $E^{(0)}_2 = 33$~meV.

\begin{figure}[htb]
  \onefigure[width=0.8\columnwidth]{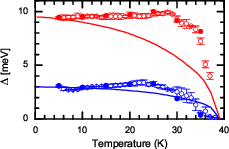}
  \caption{Superconducting gap values derived from the fine RTR measurements (open symbols) and the accurate absolute reflectivity (solid symbols). The solid lines are Suhl \etal \cite{PhysRevLett.3.552} calculations for a two band superconductor with a large interband interaction term.}
  \label{Fig4}
\end{figure}

The Mattis-Bardeen formalism assumes a weakly coupled superconductor and sets the absorption edge at twice the gap, \emph{i.e.}, $E = 2\Delta$. In strongly coupled pnictides, a full Eliashberg calculation is more accurate \cite{Charnuka2011} and the absorption onset happens at $E = E_B + 2 \Delta$, where $E_B$ is the the pairing exchange boson energy 
(the so-called Holstein process)
\cite{Akis1991}. Yet, Mattis-Bardeen remains a valid approximation for the electrodynamics of the system, as far as one takes into account the boson energy. A strong candidate for the pairing glue is the 14 meV neutron scattering magnetic resonance found in \BKFAn\ \cite{Christianson2008,PhysRevLett.108.227002}. This value leads to $T = 0$ optical gaps of $\Delta^{(0)}_1 = 3$~meV and $\Delta^{(0)}_2 = 9.5$~meV, well within the reported range for this material  \cite{EPL.83.47001,EPL.85.67002,PhysRevLett.101.257006,PhysRevB.79.012503,PhysRevB.79.174501,PhysRevLett.105.027003,PhysRevB.83.060510}. To obtain $\Delta(T)$, shown in Fig.~\ref{Fig4}, we utilized the temperature evolution of the neutron resonance peak determined for BaFe$_{1.85}$Co$_{0.15}$As$_2$ \cite{Inosov2009} 
and renormalized that curve to match the $T=0$ value measured for the K-doped material. 

Suhl \etal \cite{PhysRevLett.3.552} proposed an extension of BCS theory for multiple bands. They define intraband ($V_{11}$ and $V_{22}$) as well as interband ($V_{12}$) pairing interaction terms. When $V_{12} \ll \sqrt{V_{11} V_{22}}$ a two band material behaves almost as two independent single band superconductors, and the smaller gap (almost) closes at a temperature smaller that the macroscopic $T_c$ for the material. Conversely, when only $V_{12} \neq 0$, one finds the situation shown by the solid lines in Fig.~\ref{Fig4}, which qualitatively agrees with our data: both gaps have a similar temperature dependence and close at the same $T_c$. The measured gap opens faster than Suhl \etal\ calculations, as one would expect for strongly coupled superconductors \cite{PhysRevB.77.104510} such as pnictides. Nevertheless, the qualitative agreement indicates a strong interband pairing interaction.

Earlier, we argued that our use of two Drude peaks do not prevent the system from having more than two bands. The situation in the superconducting state is equivalent: our use of two Mattis-Bardeen terms indicates the presence of \textit{at least} two gaps. The total number of superconducting gaps may be larger, as far as these gaps have energies similar to the two shown here. 
Also note that if gap value calculations are carried out without the Holstein process one will find gaps that are larger than those measured by ARPES and STM. Nevertheless, a different gap value will not affect  the other conclusions we find in this paper.

%
%
\section{Penetration Depth}

The optical conductivity is a very accurate technique to obtain the absolute value of the penetration depth. Let us define a spectral weight function:
\begin{equation}
  S(\omega_c) = 
  	\int_{0^+}^{\omega_c} \sigma_1(\omega) \, \upd \omega \, ,
  \label{eq2}
\end{equation}
which leads to the $f$-sum rule $ S(\omega_c \rightarrow \infty) \propto \sfrac{n}{m}$, where $n$ is the total number of electrons and $m$ is the bare electron mass. This quantity is independent of any external parameter, such as the temperature, even across phase transitions. The $0^+$ limit indicates that zero frequency is not accessible directly by $\sigma_1(\omega)$. This is relevant for the superconducting state, where the infinite conductivity is represented by a $\delta(\omega)$ function. The $f$-sum rule conservation, when applied to the superconducting state, leads to the Ferrell, Glover and Tinkham (FGT) sum rule \cite{Glover1956,Tinkham1959}: spectral weight lost at finite frequencies in $\sigma_1$ below $T_c$ is transfered into the superfluid weight. One can then calculate the superfluid density from:
\begin{equation}
\Delta S(\omega_c) = S_N(\omega_c) - S_S(\omega_c) \quad,
\label{eq4}
\end{equation}
where $N$ and $S$ refer to the normal and superconducting states, respectively and $\omega_c$ is a cut-off frequency. The London penetration depth is $\lambda_L^{-2} = 4 Z_0 \Delta S$.

\begin{figure}[htb]
  \onefigure[width=0.8\columnwidth]{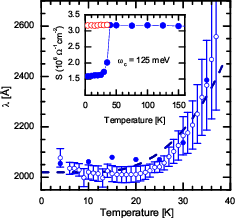}
  \caption{Penetration depth  for \BKFAn\ calculated from $\sigma_2(\omega)$ obtained from the RTR (open circles) and the accurate absolute reflectivity (solid circles). The dashed line is a BCS calculation with $\Delta^{(0)} = 9.5$ meV gap. Inset: The solid circles represent the spectral weight from Eq. \ref{eq2} ($\omega_c = 125$~meV). The open circles are obtained by adding the superfluid weight calculated from $\sigma_2(\omega)$ to $S(\omega_c)$. }
  \label{Fig5}
\end{figure}

The inset of Fig.~\ref{Fig5} shows the spectral weight for \BKFAn. The solid blue circles are a direct application of Eq.~\ref{eq2} with $\omega_c = 125$~meV. Any cut-off energy above 125~meV gives similar results. In the normal state $S$ is constant within 2\%. The decrease in $S$ below $T_c$ is the signature of the superfluid formation. $\Delta S$ is a measurement of the superfluid stiffness. 

To verify this assertion, one can use an alternative method to find the superfluid stiffness. The  $\delta(\omega)$ function in $\sigma_1$ implies, because of Kramers-Kronig relations, a $\omega^{-1}$ term in $\sigma_2$, the weight of which is also a measurement of the superfluid stiffness.  References \cite{Dordevic2002,Zimmers2004b} give details on the calculatation of $\lambda_L$ from $\sigma_2$. The open red symbols in the inset of Fig.~\ref{Fig5} are obtained by adding the superfluid weight calculated from $\sigma_2$ to the value of $S(\omega_c)$. This picture shows that the superfluid weight is built from an energy range below 125 meV, and agrees with the expectations of FGT that this energy range should correspond to a few times the superconducting gap energy.

The main panel in Fig.~\ref{Fig5} depicts the thermal evolution of the penetration depth in \BKFAn. The solid circles are calculated from the absolute reflectivity and the open circles are obtained from the RTR. 
Our absolute value for $\lambda_L$ is the same as the one found by infrared \cite{PhysRevLett.101.107004} and tunnel diode resonator \cite{PhysRevB.80.020501} measurements. It is larger than the values (100--120 \AA) obtained from critical fields measurements \cite{Ren2008}.
The dashed line is the solution of the penetration depth BCS equation assuming a strong coupling temperature dependence for a gap $\Delta^{(0)} = \Delta^{(0)}_2 = 9.5 \textrm{ meV}$. The larger gap $\Delta^{(0)}_2$, having a smaller penetration depth, dominates the value of the total penetration depth. This is a signature that the superconducting bands respond in parallel as implied by summing two Mattis-Bardeen terms in $\sigma_1(\omega)$.

%
%
\section{In-plane and out-of-plane doping}
\begin{figure}[htb]
  \onefigure[width=0.8\columnwidth]{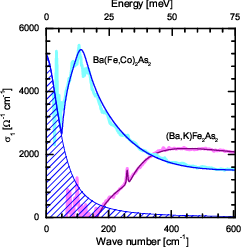}
  \caption{Optical conductivity at 4 K for \BFCAn\ (Ref.~\cite{PhysRevB.82.100506}) and \BKFAn. For the latter, the thin solid line is a fit as described in the text. For the former, besides the superconducting Mattis-Bardeen response, a Drude term (hatched area) is necessary to account for the residual superconducting state sub-gap conductivity.}
  \label{Fig6}
\end{figure}

To grasp some information on the $s_{\pm}$ symmetry, it is instructive to compare \BKFAn, where dopants are out of the FeAs planes, to \BFCAn, where dopants go in the FeAs planes. Figure~\ref{Fig6} shows the 4~K optical conductivity for both materials. In \BKFAn, $\sigma_1(\omega)$ vanishes below the smaller absorption threshold ($E^{(0)}_1$) indicating a fully open gap and the absence of unpaired carriers below $T_c$. Conversely, $\sigma_1$ for \BFCAn\ \cite{PhysRevB.82.100506} shows a large sub-gap absorption in the superconducting state. This extra absorption is described by a Drude peak (hatched area) due to unpaired carriers. 

In \BFCAn,  the in-plane scattering by Co atoms induces pair-breaking that is, at least partially, responsible for the residual low frequency absorption \cite{PhysRevB.82.100506,Fischer2010,Aguilar2010}; the residual low temperature specific heat \cite{Mu2010}; and its penetration depth \cite{Gordon2010}. In \BKFAn, K atoms stay out of the FeAs planes, therefore no interband impurity scattering is expected. Indeed, here we observe no residual low frequency $\sigma_1(\omega)$. The fact that only when dopant atoms sit on the FeAs planes does one see unpaired quasiparticles matches naturally the $s_\pm$ gap symmetry. This view also reconciles the full gap from ARPES with the V-shaped STM density of states. 

%
%
\section{Conclusions}
We report the optical conductivity of a \BKFAn\ single crystal. In the superconducting state, a flat 100\% reflectivity below 20 meV, with consequent vanishing of the optical conductivity in the same energy range, indicates a fully open superconducting gap. A Mattis-Bardeen analysis shows that, in fact, two almost isotropic gaps with different values are required to describe the optical response of our sample. The temperature dependence of the gaps is compatible with a two band superconductor with a strong interband coupling. The London penetration depth for \BKFAn\ thermal evolution is dominated by the larger gap, as expected for a superconductor with bands responding in parallel to the electromagnetic field. We find that \BKFAn\ out-of-plane K atoms do not induce pair-breaking whereas scattering by the in-plane Co atoms of \BFCAn\ depletes superconductivity. This result strongly supports an $s_\pm$ symmetry for the gap in BaFe$_2$As$_2$ superconductors.

%
%
\begin{acknowledgments}
We would like to acknowledge discussions with J. P. Carbotte and T. Timusk, and the financial support from the Science and Technology Service of the French Embassy in China. Work in Beijing was supported by the MOST and the National Science Foundation of China.
\end{acknowledgments}

%
%

\end{document}